\documentstyle[12pt,epsfig]{article}
\begin{document}
\vspace*{2cm}
\begin{center}
{\Large {   Flow with PMD: Past and Future\footnote{Invited talk at the Fourth International Conference on Physics and Astrophysics of Quark Gluon Plasma, 26-30 Nov.2001, Jaipur, India}
}}

\end{center}

\bigskip

\begin{center}

%{\small{
\noindent

Sudhir~Raniwala

%}}
\end{center}
\begin{center}
{\small{

Physics Department, University of Rajasthan, Jaipur 302004, India
}}           
\end{center}

%\newpage

\bigskip
\vskip 3cm

\begin{abstract} 

Measurements of azimuthal distribution of inclusive photons using the 
fine granularity preshower photon multiplicity detector (PMD) at CERN SPS 
are used to obtain anisotropy in the azimuthal distributions. 
These results are used to estimate the anisotropy in the neutral 
pion distributions.
The results are compared with results of charged particle data, both 
for first order and second order anisotropy. Assuming the same anisotropy
for charged and neutral pions, the anisotropy in photons is
estimated and
compared with the measured anisotropy.
The effect of neutral pion decay
on the correlation between the first order and the second order event plane 
is also discussed.  Data from PMD can also be used to estimate the 
reaction plane for studying any anisotropy in particle emission
characteristics in the ALICE experiment at the Large Hadron Collider.
In particular, we show
that using the event plane from the PMD, it will be possible to measure 
the anisotropy in J/$\psi$ absorption (if any) in the ALICE experiment.

\end{abstract}

\pagebreak

\section{Introduction}

A preshower photon multiplicity detector (PMD) was first used in the WA93
experiment to measure the inclusive photon yield and their spatial
distributions to study S+Au collisions at SPS energies \cite{nim1}.
The measurement of the azimuthal distributions of these photons 
resulted in reporting the first ever observation of anisotropy at 
these energies \cite{anisowa93}. Part of the observed anisotropy was 
due to correlations from the neutral pion decay. The observed 
anisotropy in photons could not be used to determine the anisotropy 
in neutral pions, nor could the quantitative contribution from decay 
correlations be determined at that time. However, the observed 
anisotropy was found to be greater than what was observed in a similar 
sample of VENUS events, which implicitly 
include the effect of energy momentum correlations in neutral pion decay.

There now exist methods that enable a determination of the neutral 
pion anisotropy
from a measurement of the anisotropy in azimuthal distribution of 
inclusive photons \cite{aniso_pion}. In the present paper we shall 
show the results on anisotropy in photon emission and
its centrality dependence. We shall use the method detailed in 
\cite{aniso_pion} to estimate the
anisotropy in neutral pions and compare the results with the anisotropy 
measured in the charged pions in the same pseudorapidity region. 
The effect of finite granularity of the detector(s)
and the effect of efficiency and purity of the data sample will also
be investigated.

The measures of anisotropy are the coefficients in the Fourier expansion of
the azimuthal angle distribution. The precision of these coefficients 
systematically increases for increasing
anisotropy and for increasing multiplicity. The measured 
azimuthal angle distribution of the photons in the ALICE environment
can also provide an estimate of the reaction plane of the event. The 
possibility of measuring any dependence of particle emission 
characteristics on the event 
plane direction is investigated, in particular the possible anisotropic
absorption of J/$\psi$ resonance \cite{intjpsi}.

\section{WA98 Experiment and DATA}

The preshower PMD measured the inclusive distribution 
of photons in Pb+Pb collisions on an event by event basis in the WA98 
experiment in the pseudorapidity region 2.8 to 4.2 \cite{nim2}.
The Silicon Pad 
Multiplicity Detector (SPMD) measured charged particles in the 
region 2.35 to 3.75.
The present analysis deals with the data in the region 3.25 to 3.75 
corresponding to 100\% azimuthal coverage for the two detectors in the region
of overlap. The basic features of the PMD, the minimum bias 
multiplicity distribution and the pseudorapidity distributions can be 
seen in \cite{bedanga}. Details of data reduction, efficiency and purity 
of the detected samples can be seen in \cite{nim2}.
The analysis has been carried out for the seven centrality classes.
The total number of events in every centrality class and the corresponding
percentage of cross section are shown in Table~\ref{data}.

\begin{table}[h]
\begin{center}
\caption{WA98 DATA}
\label{data}
\vskip 3mm
\begin{tabular}{ccc}
E$_T$ Cut & Percentage of Cross Section & Number of Events \\ \hline
$>$ 347.6 & ~0 - ~5 \% & 68 K \\
298.6 - 347.6 & ~5 - 10 \% & 75 K \\
225.5 - 298.6 & 10 - 20 \% & 54 K \\
170.2 - 225.5 & 20 - 30 \% & 18 K \\
124.4 - 170.2 & 30 - 40 \% & 19 K \\
89.9 - 124.3  & 40 - 50 \% & 18 K \\
40.0 - 89.9  & 50 - 80 \% & 37 K \\
\end{tabular}
\end{center}
\end{table}

\section{Analysis and Results}

For every event, the event planes of both orders are determined using the
relation
 
\begin{eqnarray}
      \psi_n= \frac{1}{n}\left(\tan^{-1} \frac{\Sigma w_i \sin n\phi_i}
        {\Sigma w_i \cos n\phi_i}\right)
\label{event}
\end{eqnarray}

\begin{figure}[htbp] 
\epsfxsize=12cm
%\centerline{\epsfbox{psi2corr_pmd.eps}}
\centerline{\epsfbox{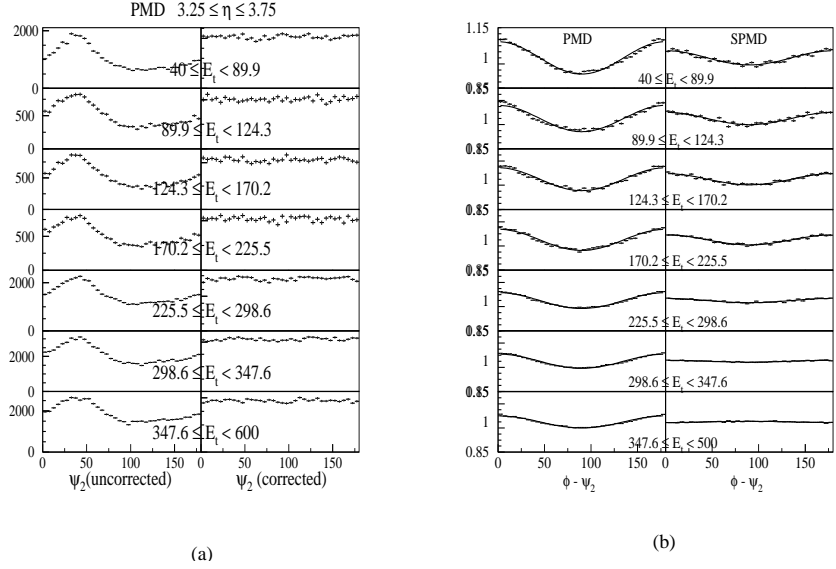}}
\caption{ (a) $\psi_2$ distribution of $\gamma$-like hits for various centralities
before and after the acceptance correction factors (b) Azimuthal angle 
distribution of particles with respect to
the second order (corrected) event plane angles. First column is for 
$\gamma$-like in the PMD and the second column is for the charged particles
in the SPMD }
\label{flatpsi2}
\end{figure}

\noindent The $\phi_i$ are the azimuthal angles of the emitted 
particles and the 
w$_i$ are the weight factors. For the azimuthal distribution of the 
number of particles, these weight factors are all equal to one. 
The data have been corrected for acceptance. The azimuthal distribution 
of the inclusive photons (and of charged particles) is plotted for 
every centrality bin and the 
inverse of 
occupancies have been taken as weights in obtaining the event plane 
angles in the equation above. This 
has been done for both PMD and SPMD for all centralities. A typical 
distribution for the second order event plane angles is shown in 
Fig.~\ref{flatpsi2}a for the $\gamma$-like particles recorded in the PMD, 
both before and after the acceptance correction. $\gamma$-like particles 
include
the inclusive photons and some charged particles which deposit an energy
comparable to energy deposited by a photon \cite{nim2}.The flatness of 
the $\psi_2$ distribution shows that the detector acceptance effects 
have been corrected for. 

The distribution of the azimuthal angles of the emitted particles 
with respect to the event planes enables a determination of the 
coefficients v$_1$ and v$_2$ characterising directed and elliptic flow.
Both these coefficients are determined in more than one way, 
differing in details and is a good recipe for consistency checks.

We determine 
\begin{enumerate}
\item $v_1'$ = $\langle \cos (\phi - \psi_1^{est}) \rangle $ :  a
measure of the directed flow in the direction of $\psi^{est}_1$, 
where the average is over all particles of all events. 
\item $v_2'$ = $\langle \cos 2(\phi - \psi_2^{est}) \rangle $ : a 
        measure of the ellipticity about $\psi_2^{est}$
where the average is again over all particles of all the events. 
\end{enumerate}

The distribution of the azimuthal angles of the particles with respect to
the event plane is  shown in Fig.~\ref{flatpsi2}b for the second order, 
both for PMD and for SPMD, for all centralities. To avoid auto 
correlations, the event plane 
has been obtained from all the particles excluding the particle in question.
The v$_n'$ values are then determined by fitting this distribution to 
\begin{eqnarray}
      r(\phi)= \frac{1}{2\pi}\left[1 + 2v_1' \cos(\phi -\psi_1) +
 2v_2' \cos 2(\phi-\psi_2)\right]
%\label{aniso}
\end{eqnarray}

Distribution of particle azimuthal angles is also obtained with respect 
to the acceptance uncorrected event plane angle. Similar distribution
is obtained for mixed events. The effect of detector anisotropy cancels
in the ratio of the two  
distributions and is used to determine the v$_n$' values. 
The v$_n'$ values obtained this way are the same as obtained for the case 
by fitting the distribution to Fig.~\ref{flatpsi2}b.
The prime on the v$_n$ indicates that these are projected on the estimated
reaction plane and not the true reaction plane and therefore need a 
correction factor.

\subsection{Event Plane Resolution Correction}
Since the estimated event plane fluctuates about the actual event plane 
due to finite particle multiplicity, the average correction 
factor for the event plane determination, termed the resolution correction 
factor (RCF) is determined. Experimentally, RCF is obtained using 
the  subevent method
        described in \cite{voloshin}. 
         Here every event is  divided randomly into two 
         subevents  of equal multiplicity and the angle $\psi_n$ 
is determined for each subevent.
This enables a determination 
        of a parameter $\chi_n$ directly from the experimental data
using the relation \cite{voloshin,ollie2}:
\begin{eqnarray}
        \frac{N_{events} ( n | {\psi_n^a - \psi_n^b}| > \pi/2)}{N_{total}} 
= \frac{e^{-\frac{\chi_n^2}{4}}}{2}
\label{chim}
\end{eqnarray}
\noindent where $N_{total}$  denotes the total number of
events, $\psi_n^a$, $\psi_n^b$ are the estimated 
angles of the two subevents (labelled
$a$ and $b$) and the numerator on the left denotes the
 number of events having the angle between subevents greater
than $\pi/2n$. 
The parameter $\chi_n$ 
  is then used to determine RCF according to the relation given in the Ref.
\cite{voloshin}.

\begin{figure}[htbp] 
\epsfxsize=8cm
\centerline{\epsfbox{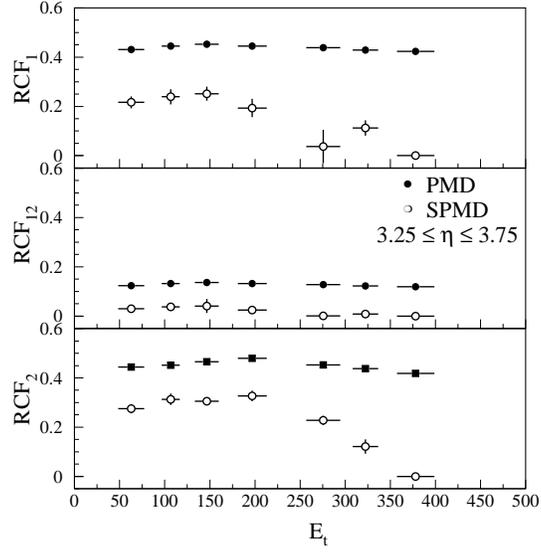}}
\caption{ The event plane correction factors for charged particles and 
for $\gamma$-like for the different orders of anisotropy 
}
\label{rcf}
\end{figure}

For the data sample used in the present analysis,
the resolution correction factors are shown in Fig.~\ref{rcf}. 
RCF$_n$ is the correction factor for anisotropy of the order n and
RCF$_{12}$ is the correction factor required when one determines the 
second order anisotropy with respect to the event plane of the first order.
Smaller values of these factors 
introduce larger errors on the estimated values of the anisotropy. 

\subsection{Anisotropy Values}

\begin{figure}[htbp] 
\epsfxsize=8cm
\centerline{\epsfbox{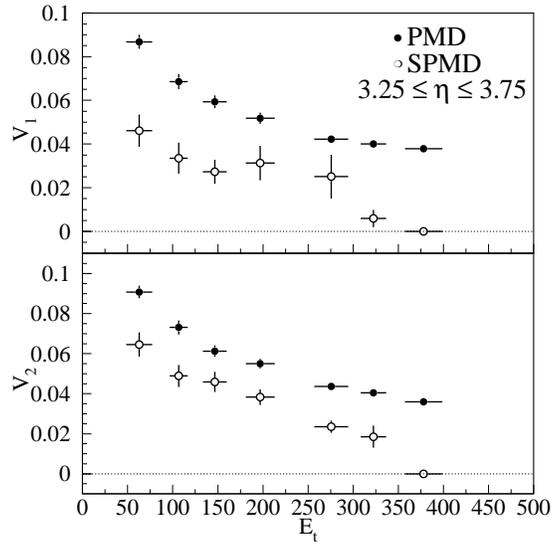}}
\caption{ v$_n$ for different centralities for $\gamma$-like and for charged
particles }
\label{vnvalues}
\end{figure}

The resultant values of v$_n$ for the PMD and for the SPMD as a function
of centrality are shown in Fig.~\ref{vnvalues}.
The v$_n$ values are also obtained by dividing $\chi$ obtained in 
eqn.~\ref{chim}
above by the fluctuation in the multiplicity in that centrality bin.

The anisotropy parameters for the $\gamma$-like sample are systematically 
greater than the corresponding parameters for the charged particles 
and the anisotropy parameters for both of them decrease monotonically 
with increasing centrality.

\subsection{Decay Effect} 
The decay of the neutral pions into photons introduces correlations 
because of energy momentum conservation, causing an apparent anisotropy 
in the photons which is different from the actual anisotropy present in the 
pions. The process of decay also dilutes the initial correlation 
present in the neutral pions. The resultant effect of these two competent 
effects scales with the experimentally measured quantity
$\chi$ and enables a determination of the neutral pion anisotropy
using the relation ~\ref{pionani} as obtained from reference 
\cite{aniso_pion}

\begin{eqnarray}
        \frac{v (\gamma)}{v^{in}(\pi^0)} = \frac{a}{(\chi-b)^2} + c
\label{pionani}
\end{eqnarray}

\noindent The values of the constants in this relation depend weakly 
on the details
of the realistic kinematic distributions of the charged pions. For the 
present work the values are taken from the reference cited above. For first 
and second order, the values of (a,b,c) are respectively (0.075, 0.473, 0.801)
and (0.031, 0.286, 0.619).

\begin{figure}[htbp] 
\epsfxsize=8cm
\centerline{\epsfbox{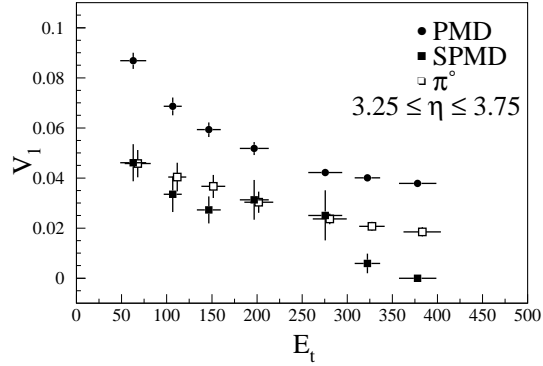}}
\caption{ v$_1$ for different centralities for $\gamma$-like as obtained 
from the PMD and for charged particles from SPMD. Using v$_1$ from PMD 
and the scaling relation, v$_1$ for charged particles is estimated.
}
\label{scaled}
\end{figure}

Using the measured values of v$_1$ in the PMD, and using 
relation ~\ref{pionani}, the v$_1$ values for neutral pions are obtained,
enabling a comparison with the corresponding values for charged particles.  
The anisotropy
values for charged particles, for $\gamma$-like in the PMD and the deduced
anisotropy for neutral pions are all shown in Fig.~\ref{scaled}. Where 
this agreement is quite good, there are limitations. Certain effects not
incorporated are: 
the effect of finite granularity on the measured values of 
anisotropy, the effect of charged particle contamination in the $\gamma$-like
sample, the effect of protons on the SPMD and other effects.

\subsection{Granularity,Efficiency and Other Effects}
The PMD has very fine azimuthal granularity, and the finite granularity
does not have any effect on the estimated anisotropy values. The azimuthal
bin size of the SPMD is 2 degrees. 
This not-so-fine granularity results in a measured charged particle
yield which is less than the actual multiplicity. Since the number 
of 'hits' and their distribution in SPMD are used, the 'coarse' granularity
also has an effect of diluting the actual azimuthal
correlation. Events are simulated with initial anisotropy values and 
multiplicity. The  
granularity of the SPMD is incorporated before analysing the data for
anisotropy. The initial conditions are varied to obtain the 
experimentally
measured values of the anisotropy parameters for charged particles. Using
this anisotropy value along with the corresponding multiplicity for 
neutral pions, followed by neutral pion decay kinematics enable to
estimate the anisotropy in photons. These estimates of anisotropy are 
compared to those measured in the PMD, for both orders.

\begin{figure}[htbp] 
\epsfxsize=8cm
\centerline{\epsfbox{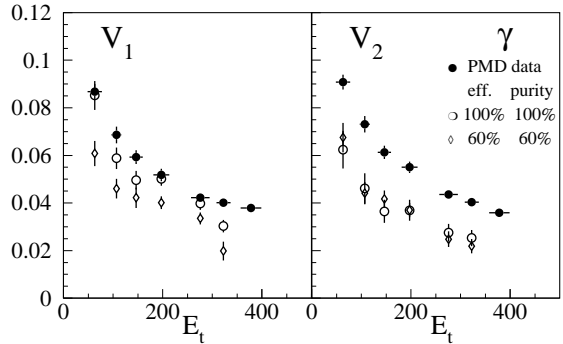}}
\caption{ v$_n$ for different centralities for $\gamma$-like as obtained 
from the PMD. Using v$_n$ values of charged particles and decay
kinematics, expected values for photon samples are also plotted 
for two different combinations of efficiency and purity
}
\label{result}
\end{figure}

The anisotropy measured in the PMD are observed to be larger than what 
is obtained by the method described above. 
Further, 60\% of the $\gamma$-like sample is assumed to be decay photons 
from these neutral pions and 40\% of the sample is contaminants.
We assume that these contaminants carry the same anisotropy as those 
of the charged particles. We see
broad agreement for the first order (v$_1$) but for the second
order, v$_2$ observed in PMD is systematically larger than what can be 
obtained by using the values of v$_2$ for charged particles and using
decay kinematics. The results are shown in Fig.~\ref{result}.
These results are preliminary, and at present there are no known 
explanations for this excess.

\subsection{Event Plane from PMD}
\begin{figure}[h] 
\epsfxsize=8cm
\centerline{\epsfbox{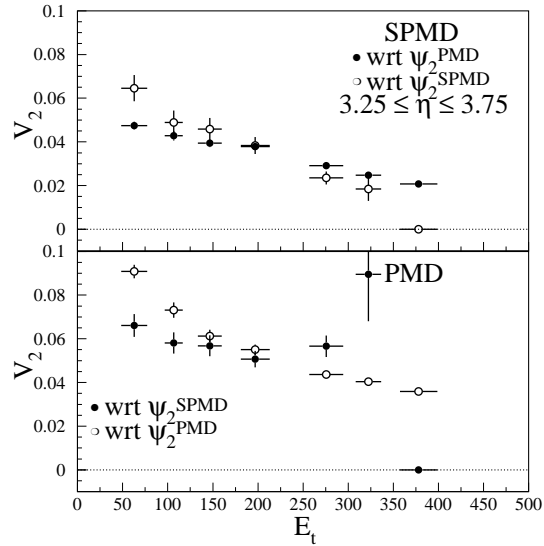}}
\caption{ v$_n$ for different centralities for $\gamma$-like with respect to 
the event plane obtained in the spmd and for charged particles with 
respect to event plane obtained in the pmd
}
\label{pmdspmd}
\end{figure}

Independent of asymmetries in particle emission as studied above,
it is possible to  determine an event plane from the PMD and investigate 
any possible asymmetries in charged particle emission, with respect
to the event plane. By construction, the asymmetry coefficients 
(Fourier coefficients) obtained earlier will result in the maximum
value. The event
plane obtained from the PMD may have a little bias as compared to the
event plane for the charged particles due
to decay.  Event planes estimated for different particle species in
different rapidity regions could all be different, beyond fluctuations,
the difference arising due to different final state 
effect on the particles while they stream from the point of freezeout
to the detectors.  It is then expected that
v$_2$ for charged particles with respect to  event plane of the 
PMD would be less 
than v$_2$ as measured using the event plane from the SPMD. The same 
result is expected if for the estimate of  
v$_2$ for $\gamma$-like using the event plane of the SPMD. This is observed
and is shown in Fig.~\ref{pmdspmd}.
This demonstrates that the event plane estimated from the PMD can be 
used to study emission properties of any particle species, in any other 
acceptance region.
%\begin{figure}[htbp] 

\subsection{Correlation Between Event Planes of Different Orders}
A priori, one expects that the event planes of both the orders are 
either the same, or perpendicular to each other, depending upon the 
speed of sound in the colliding matter and the colliding energy of 
the nuclei.  The estimates
of these event planes are affected with fluctuations. 

\begin{figure}[h] 
\epsfxsize=10cm
\centerline{\epsfbox{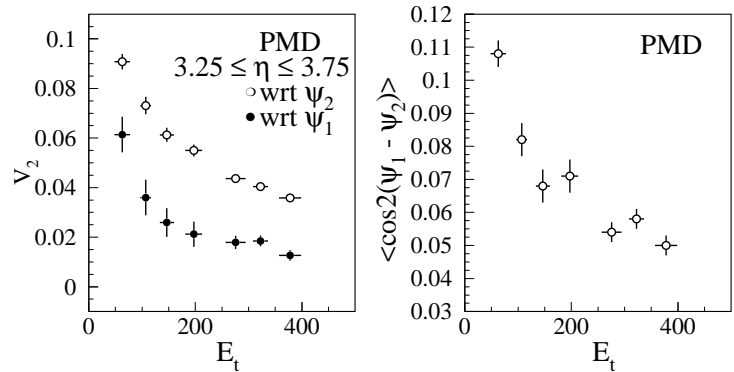}}
\caption{ (a) v$_2$ for different centralities for $\gamma$-like with 
respect to 
the event plane of first order and event plane of second order. (b) The correlation between the first order and the second order event planes in the PMD
}
\label{correla}
\end{figure}

However, correcting for these fluctuations
should result in the same numerical value of
v$_2$ determined from event plane of either order. The data from the 
PMD suggests that this is not the case, where v$_2$ determined from the 
first order event plane from the PMD is systematically lower, in addition 
to having larger error bars. The results are shown in Fig.~\ref{correla}a.
This indicates that one has to 
look for possible correlations (or lack of correlations) between the 
event plane of the 
two orders. 
The 
event planes of the two orders may not be correlated due to effects 
such as initial state density fluctuations. The absence of this 
correlation can also arise if the event plane is estimated using particles 
which are products of decay. Then decay kinematics affects the event
planes of different orders differently.     
To look for correlations between the event plane angles,
the quantity  
$\langle \cos 2 (\psi_1 - \psi_2) \rangle $ 
is deduced, where the 
average is over the event samples in question. This is plotted 
as a function
of centrality and is shown in Fig.~\ref{correla}b. This behaviour is
explicable by the decay effect. Starting with the same direction for 
the first and the second order event plane, and the
different centralities  characterised by  different
mean multiplicities, a simulation which includes the neutral pion
decay is able to reproduce the observed correlation between the event
planes of the two orders. This
shows that the 'primordial'
event plane is affected by the decay of the particles, and the 
quantitative effect is different for the event planes of the two orders.
This results into different values of v$_2$ as determined from the event 
planes of the two orders. 
%\begin{figure}[htbp] 

\section{PMD in the ALICE Experiment}

A logical extension of the results obtained from PMD in WA98 experiment 
was to investigate whether the 
PMD could measure azimuthal anisotropies in photon emission in the 
ALICE experiment. The results of these investigations are detailed in 
reference ~\cite{pmdtdr}. Further, can the PMD  provide an event plane 
estimate in the ALICE environment
and if this event plane may be used for studying asymmetries in
particle emission. This was explored using fast simulation techniques,
is detailed in reference ~\cite{intjpsi}, and we discuss it here. 

The Physics motivation comes from the suggestion that the 
absorption of J/$\psi$ by comovers is predicted to be azimuthally asymmetric
as compared to an isotropic Glauber absorption in the forward 
region \cite{heiselberg}. The asymmetry has been
predicted to be about 5 - 20 \% depending upon the centrality, 
size and the energy of the colliding nuclei. 
Any observation
of this asymmetry in the yield of J/$\psi$ will enable a  
determination of the relative contribution of the two absorption
processes, essential to make an estimate of the actual
suppression of the resonance from the possible Quark Gluon Plasma.

The ALICE experiment proposes to measure the muons in the DiMuon 
Spectrometer D$\mu$S. The PMD is designed to take data at
the highest rate possible in ALICE, combining the trigger
rates for TPC and D$\mu$S. This makes it a specially
attractive detector to provide the event plane necessary
for the study of anisotropy in J/$\psi$ absorption.
The possibility of measuring any such 
anisotropy in the J/$\psi$ emission with respect to the event 
plane measured from the 
PMD is investigated.

\subsection{Methodology}

The data for the investigation is simulated. The 
transverse momentum and pseudorapidity distribution of the pions 
and the kaons are generated by parametrising
HIJING events. These are generated in the $\eta$ range 0 to 6 
and p$_T$ range 0 to 8 GeV/c. The central events 
correspond to a maximum charged particle pseudorapidity
density of about 6000. A branching ratio of 11.7 mb is assumed for 
J/$\psi$ decaying
into two muons. The yield of muons from other sources 
is not known a priori, and therefore this number is a
free parameter. The muons are added in the acceptance region
of the D$\mu$S and their number is proportional to the charged 
particle multiplicity in the event. 
The transverse momentum distribution of these muons was motivated
by the distribution of the average number of muons per event greater than
a particular p$_T$, as in reference \cite{muontdr}.

The azimuthal angles of all the pions and kaons are generated with a 
certain initial anisotropy. The azimuthal angles of the J/$\psi$s are 
generated assuming the absorption is maximum along the event plane.
This produces a maximum yield of the J/$\psi$ in a direction normal to 
the event plane determined from the charged particles or from the photons.

\subsection{Detector Effects}

The hits on the PMD arise from both photons and charged hadrons. After
filtering hadrons using appropriate algorithms, the remaining hits,
called ``$\gamma$-like'', contain 60\%  decay photons. The azimuthal
distributions of these photons reflect the anisotropy present in parent
$\pi^0$. Of the 40\% contaminants, half are assumed to be initial charged 
pions,
which have the same anisotropy as the neutral pions, and the remaining half
are  secondaries likely to be randomly distributed. The generated events
contain $\gamma$-like hits having suitable mix of charged particles and decay
photons according to the above proportion.

The effect of upstream material, which results in scattering of
 the incident particles, has been investigated for the coverage  
of the PMD in 
Ref.~\cite{pmdtdr}. This effect is incorporated in the 
fast simulation by allowing all particles to
change direction so that the final  $\eta-\phi$ differs
from the original values by an amount $\delta \eta$ (=0.1,0.2) ,$ \delta \phi 
(= 10^\circ,20 ^\circ) $ given
by a Gaussian distribution. The smaller values are termed as moderate
scattering and the larger values correspond to what we term as large 
scattering.

Muon track reconstruction efficiency is assumed to be 0.9 in the D$\mu$S. 
The efficiency has been varied between 0.8 to 1.0 and its 
effect on the sensitivity of the analysis is also studied. 
The mass resolution of the J/$\psi$ is considered to be 60 MeV.

\subsection{Analysis and Results}

\begin{figure}[htbp] 
%\begin{figure}[HB] 
\epsfxsize=8cm
\centerline{\epsfbox{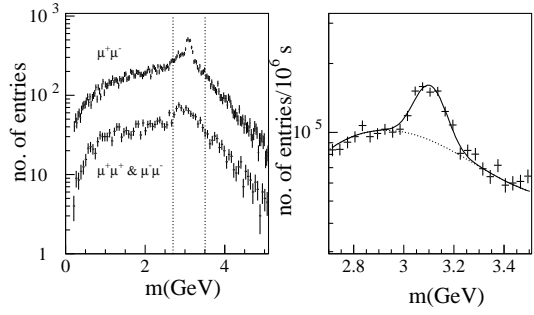}}
\caption{ 
The invariant mass spectra of the opposite sign and the like sign dimuons.
The fits to the background and the signal. 
}
\label{invmass}
\end{figure}
\begin{figure}[htbp] 
%\begin{figure}[H] 
\epsfxsize=8cm
\centerline{\epsfbox{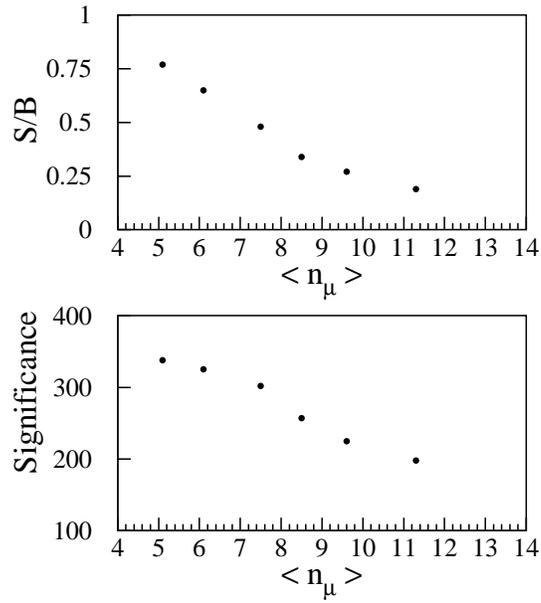}}
\caption{ 
S/B ratio and Significance as a function of number of added muons in the D$\mu$S}
\label{signi}
\end{figure}

The invariant mass spectra of opposite sign dimuons is constructed using 
muons with p$_T > $ 1 GeV and the two muons following p$_T^{min}$ + 
0.5 p$_T^{max} >$ 2 GeV. To extract the resonance yield, the invariant
mass spectra of the like sign dimuons is also obtained with the same 
kinematic cuts. The distribution is then fitted to a polynomial in the region
of the resonance mass. The opposite sign dimuon spectra is then fitted to
this polynomial and a Gaussian and the yield of the resonance is 
determined as shown in Fig.~\ref{invmass}

 Since the number of added muons is a free parameter in the  
simulation, the results for the S/B ratio and the significance of the 
data sample vary
with the number of added muons. 
The results are shown in Fig.~\ref{signi}.

\begin{figure}[htbp] 
%\begin{figure}[hb] 
\epsfxsize=11cm
\centerline{\epsfbox{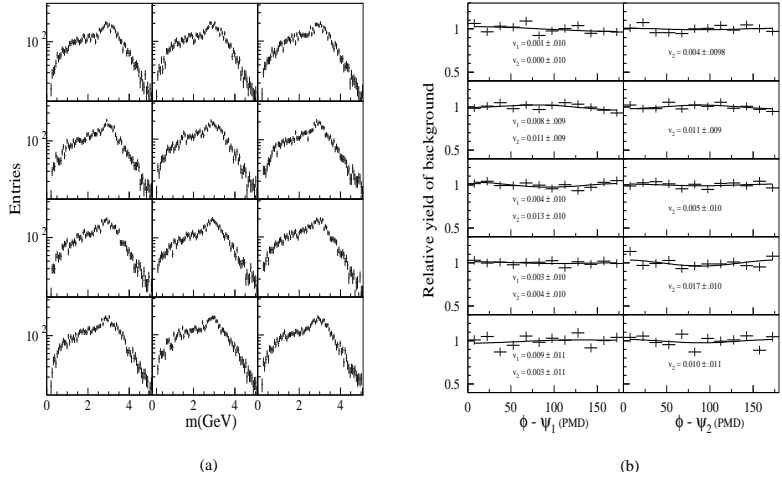}}
\caption{ 
(a) Invariant mass spectra of the like sign muons in the different azimuthal
bins with respect to the second order event plane from the PMD. (b) Fourier
analysis of background yield with respect to the event planes of first and 
second order in the PMD. 
}
\label{bkganiso}
\end{figure}

The event planes of both orders are obtained for every event and the 
resolution correction factor to correct for fluctuations of the event plane
due to finite particle multiplicity.
To look for any possible anisotropy in the background spectra, 
the invariant mass distribution of the like sign muons is plotted 
in different
azimuthal bins with respect to the event plane, and is shown in 
Fig.~\ref{bkganiso}a. The yield in different azimuthal regions in a fixed 
mass bin is used for obtaining 
the anisotropy parameters. This is repeated for different (overlapping) 
mass bins and the anisotropy parameters determined.  
There is no measurable anisotropy in the background spectra in the region
around the resonance mass, as is shown in Fig.~\ref{bkganiso}b.

\begin{figure}[htbp] 
%\begin{figure}[H4]  
\epsfxsize=6cm
\centerline{\epsfbox{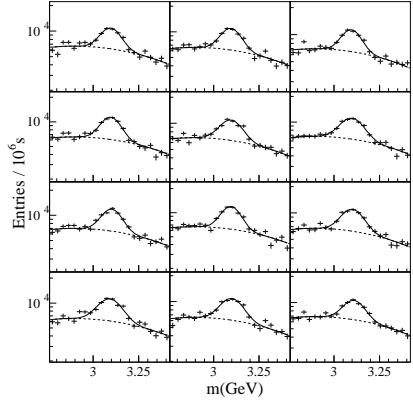}}
\caption{ 
Invariant mass spectra of opposite sign muons in the different azimuthal
bins with respect to the second order event plane from the PMD. 
}
\label{jpsiyield}
\end{figure}

The invariant mass spectra in different azimuthal regions 
is shown in Fig.\ref{jpsiyield} and the J/$\psi$ yield  
is then analysed to extract the anisotropy
components that describe the azimuthal distribution.
This has been repeated for the variety of cases where the particles
incident on the PMD remain unscattered by the upstream material, undergo
moderate scattering or undergo large scattering. 

\begin{figure}[hb] 
\epsfxsize=7cm
\centerline{\epsfbox{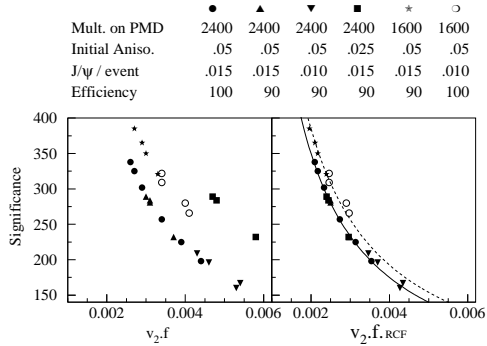}}
\caption{ (a) Significance as a function of $v_2 \cdot f$ where $v_2$ is 
the anisotropy to be probed with a fractional error $f$.(b) Same as (a) 
without the error due to event plane determination. The fitted
curve corresponds to the {\bf minimum} Significance required.
}
\label{scale}
\end{figure}

The observation of anisotropy in J/$\psi$ emission depends on several 
factors, e.g., background 
subtraction in the invariant mass spectra, increase in background due to charm
 and beauty decays, track reconstruction efficiency for muons in the DiMuon
Spectrometer and the yield of J/$\psi$ from higher resonances. 
A systematic 
study of variation in these parameters suggests that the {\it significance }
of the data (and  hence the number of events required) follows a simple 
relation with the anisotropy values to be studied and the associated accuracy,
and is shown in Fig.\ref{scale}. The fitted curve is {\it Significance = 0.7/v$_2$$\cdot f$} where v$_2$ is the value of anisotropy to be probed and f is the 
fractional error.
%\begin{figure}[htbp] 

For a given S/B ratio and muon track reconstruction efficiency, this 
significance can be translated into the number of events required to enable 
such a determination.  

%\pagebreak

\section{Summary}

The photon multiplicity detector has been successfully implemented in the 
WA93 and WA98 experiments. The first evidence of azimuthal anisotropy at 
the SPS energies was reported using measurements from the PMD. Subsequently,
PMD has obtained systematic dependence of anisotropy coefficients on 
centrality and the results have been compared with corresponding
results from charged particles. We observe that the elliptic flow 
observed in the photons is greater than the one expected by assuming 
same flow for neutral and charged particles. The anisotropy analysis 
techniques are such that the error on the estimated anisotropy coefficients 
are likely to decrease with increasing multiplicity and increasing anisotropy. 
The PMD at STAR and at the ALICE experiment will be able to provide more
precise estimates.We have also shown that the event plane from the PMD 
can be used to study 
the anisotropic characteristics of particle emission in the ALICE experiment
at the LHC.

\section{Acknowledgements}

I wish to thank all the collaborators for help and discussion at various stages
of this work and to Dr.Y.P.Viyogi for a critical reading of the manuscript. 
Part of the work was done at CERN under the CERN-ASIA programme and the CERN
Associateship programme. Financial support from the Department of Science 
and Technology of the Govt.of India is gratefully acknowledged.

\end{document}